\documentstyle [12pt]{article}

\begin{document}
\begin{titlepage}

\begin{center}
{\Large\bf{ One-loop dimensional reduction of 
the linear $\sigma$ model. } }\\ 

\vspace{.3in}{\large\em 
A.P.C.Malbouisson, M.B.Silva-Neto and N.F.Svaiter*}\\
 Centro Brasileiro de Pesquisas Fisicas-CBPF\\
 Rua Dr.Xavier Sigaud 150, Rio de Janeiro, RJ 22290-180 Brazil

\end{center}

\begin{abstract}

We perform the dimensional reduction of the linear $\sigma$ 
model at one-loop level. The effective potential of the reduced 
theory obtained from 
the integration over the nonzero Matsubara frequencies 
is exhibited. Thermal mass and coupling constant 
renormalization constants are given, as well as the thermal
 renormalization 
group equation which controls the dependence of the 
counterterms on the temperature. We also recover, for the reduced 
theory, the vacuum unstability of the model for large N.
\end{abstract}

Pacs numbers: 11.10.Ef, 11.10.Gh 

\end{titlepage}

\newpage\baselineskip .37in

\section {Introduction}

The $O(N)$ invariant linear $\sigma$ model is particularly
 interesting in the 
study of high-density matter. For instance, when $N=4$ it may be
 taken as a model for
the low energy dynamics of QCD, and also it forms one of the basic 
theoretical tools for the description 
of the pion-condensed state of neutron stars matter. 
At low temperatures, the model presents a spontaneous 
breakdown of the $O(N)$ rotational invariance. One though expects a 
second order phase transition at some temperature, since the linear
 $\sigma$ model is in the same class of universality of the 
classical $N$ component Heisenberg ferromagnet. 
In the context of QCD, high temperature expansion of static 
(i.e. having external legs corresponding to the zero Matsubara mode) 
two-point functions have been obtained by Gale 
and Kapusta \cite{Gale}. 
However, it is, in general, very difficult to carry out 
these calculations 
beyond the leading order in the loopwise expansion. On general grounds,
 in  order to circunvect those computational difficulties we relieve,
 among others, two different methods of studying the high-temperature 
properties of a quantum 
field theory. The first one is based on a resumed perturbation theory 
where the 
introduction of a generalized effective action $\Gamma(\phi)$, which 
depends not only on the field $\phi$ but also on a composite operator
 $G(x,y)$, allows one to sum up an infinite set of 
higher order diagrams \cite{Pi}. The second one is the so called 
dimensional reduction \cite{Appelquist-Pisarski}, which is based in 
the Appelquist-Carazzone 
decoupling theorem \cite{Appelquist-Carazone}. This method allows 
one to relate very hot thermal correlation functions in a 
$D$-dimensional space to zero temperature correlation functions 
in a space of dimension $D-1$, giving a "dimensionally 
reduced" effective theory. In other words, hot field theories in 
$D=d+1$ dimensions are related 
to zero temperature effective field theories, in $d$ dimensions, 
at the price that the renormalized parameters of this new theory
 have a dependence on the temperature \cite{Jourjine}\cite{Landsman}.
 Those must be considered 
at the limit of infinite temperature as new bare parameters subject
 to a 
second renormalization procedure so that the reduced theory makes
 sense.

The dynamical study of thermal field theories is based on the 
computation of the partition function $Z=tr e^{(-\beta H)}$,
where $H$ is the Hamiltonian and $\beta^{-1}$ is the temperature.  
The fact that in QCD and in other models of quantum field theory at
 finite temperature we have dramatic long range 
correlations when any of the boson masses approaches zero is one
of the most important yet unsolved problems in thermal quantum field
theory. For a recent discussion on this subject see for instance
 \cite{Braaten}.
In general, for thermal, or even non-thermal,
field theories there is not a procedure equivalent to the
Bloch-Nordsieck mechanism in QED, for the computation of the physically
relevant quantities. In QED, this procedure changes the definition
of the asymptotic states in order to include all soft photons, giving  
infrared finite coss-sections. In general quantum field theories
the infrared divergences causes 
the breakdown of perturbative calculations at some loop order and one 
would be left only with the hope of finding non-perturbative
techniques to deal with theories like, for example, QCD. As far as we
 know, we are still far from having a satisfactory answer to this
 problem. 
Even if we use lattice simulations to compute correlation 
functions for static observables 
in hot QCD, we see that the computational resources that 
are required increase rapidly with $\beta^{-1}$. It is
perhaps interesting to recall that in the case of zero 
temperature QCD, the non-cancellation of 
infrared divergences is expected to be closely related to the
 confinement mechanism \cite{Altes-Rafael} \cite{Frenkel-Taylor} 
\cite{Frenkel-Frenkel}.
On the other side, for finite temperature QCD, it is expected that,
 at very 
high temperatures, QCD undergoes a color deconfinement 
phase transition where hadrons would melt in a quark-gluon plasma.
 The connection between these two aspects,
the infrared behavior of QCD and the deconfining phase transition
 at high
temperatures has been investigated, by many authors, since the
 pioneering
work of Kalashnikov \cite{Kalashinikov}. 
An interesting question would be to investigate whether this phase
transition (the existence of some critical temperature) appears for
hot QCD in the context of Matsubara formalism. Then it would be natural
 to expect some features of the $D-1$ dimensional reduced QCD to be 
those of a quark-gluon plasma in this space dimension as well. 

At this point dimensional reduction is considered as a nice tool to 
investigate the above problems. First of all, the leading infrared 
behavior of four dimensional QCD at high temperature is governed by 
the static (zero Matsubara frequency) sector, obtained by integrating
 out the nonstatic modes, leaving an effective three dimensional field
 theory. Indeed, the introduction of temperature give us propagators of
 the kind
\begin{equation}
\Delta({\bf p},\omega_{n})=\frac{1}{\omega_{n}^{2}+{\bf p}^{2}+m^{2}},
\label{prop}
\end{equation}
where the $\omega_{n}'s$ are the well known Matsubara frequencies
$\omega_{n}=2\pi n/ \beta$ for bosons and
$\omega_{n}=2\pi (n+\frac{1}{2})/ \beta$ for fermions. 
Infrared divergences of massless
$(m=0)$ thermal field theories arise solely from the blowing up of
the $(\omega_{n}=0)$ terms in the sum over boson energies as 
${\bf p}\rightarrow 0$. In this sense we will construct an effective
three-dimensional theory with purely $\omega_{n}=0$ terms and all
$\omega_{n} \neq 0$ absorbed into the physical parameters of the
effective hamiltonian. The only modes that propagates over
distances larger than $\beta$ are the bosonic $n=0$ modes, and they
will be the only responsible for infrared divergences of the theory.
On the other side, in the high temperature limit, new interaction terms
 will arise, all with temperature dependent coeficients. 
Although it is not known, in general, very much about the exact form of
 the 
temperature dependence of the parameters in the effective
theory, this dependence does not have any connection with the infrared
 behavior of the model and the parameters are expected to be analytical
 functions of the temperature. Fortunately, this is all the
 information
we need for our purposes. There are other ways of obtaining a 
dimensionally reduced theory without the necessity of integrating
out degrees of freedom \cite{Braaten-Nieto}, however, to the order
we are concerned, all those methods give the same answers.

Concerning the color deconfinement there are still unanswered 
questions. Does QCD have a finite temperature phase transition?
 What is its order and what would be its critical temperature? 
As pointed out, for instance, by 
Kalashnikov, the quark-gluon plasma should exist in reason of the 
smallness of the effective interaction parameter at suficiently high 
temperatures, which allows to use perturbative tools to study some
 properties 
of the plasma. The absence of confinement in QCD 
as $T\rightarrow\infty$ 
has been proved by Polyakov \cite{Pol} and Susskind \cite{Suss},
 which 
implies that a quark-gluon plasma must be formed when the temperature 
is large enough. Clearly the deconfining phase transition must 
exist in order to explain the rigorous results of Polyakov and
 Susskind. 
Phenomenologically such a phase transition is expected and even 
in a rough sense "exhibited" (see for instance \cite{Joao}). However 
from a theoretical point of view it has not yet been possible to 
put both the quark-gluon plasma and the condensed hadronic phase within 
a single framework. Condensation of the quark-gluon plasma 
into hadronic matter, which must take place at temperatures 
about $200 Mev$, is not yet clear and we have at our disposal only 
qualitative or pure phenomenological considerations to explain 
such a phenomenon. We believe that dimensional reduction 
is a useful tool for understanding the very high temperature regime 
of quantum fields. In particular we expect from the study of simple 
models as the one treated in this paper, to learn about the mechanism 
of passing from low to very high temperature states, with the hope 
that this learning could be later applied in the study of more 
realistic models as QCD.

Our choice of the linear $\sigma$
model \cite{Gell-Mann-Levy} was based in the fact that, as
previously said, when $N=4$ we have a model for the low-energy 
dynamics of QCD \cite{Kapusta}. At the same time, the 
model has other features we are interested in, like spontaneous 
breakdown 
\cite{Dolan-Jackiw} of 
$O(N)$ rotational invariance by the vacuum and a symmetry-restoring
 phase at
finite temperature.

The paper is organized as follows. 
In section II we briefly discuss the linear $\sigma$ model at finite 
temperature. In section III, the effective potential of the reduced
 theory is
 obtained as well as the counterterms that allow dimensional reduction.
Conclusions are given in section IV .
In this paper 
we use $\frac{h}{2\pi}=c=1$.

\section{The linear $\sigma$ model at finite temperature}

We are interested in studying the behavior of a model of an
 $N$-component real massive 
scalar field $\phi=\{ \phi^{a}(x) \}$ with the usual 
$\lambda (\phi^{a}\phi^{a})^{2}$ self-interaction, defined in a static
spacetime in the limit of high temperature. Since the manifold is
 static, 
there is a global timelike 
Killing vector field orthogonal to the spacelike sections. 
Due to this fact, energy and thermal equilibrium have a precise
 meaning. 

The partition function can be represented as a
functional integral over fields $\phi(t,{\bf x})$ defined on a
 time interval
$0$ to $-i\beta$. It is usual to change variables from $t$ to
 imaginary time
$\tau=it$, which implies that the 
partition function is given in terms of the Euclidean 
functional integral
\begin{equation}
Z=\int {\cal D}\phi e^{-S_{E}(\phi(\tau,{\bf x}))}, 
\label{generator}
\end{equation}
where
\begin{equation}
S_{E}=
\int_{0}^{\beta}d\tau\int d^{3}{\bf x}{\cal L}(\phi(\tau,{\bf x})), 
\label{action}
\end{equation}
is the classical action.
The Lagrangean density  of the linear $\sigma$ model
is given by
\begin{equation}
{\cal L}(\phi^{a}(\tau,{\bf x}))=
\frac{1}{2}(\partial_{\mu}\phi^{a})^{2}-
\frac{1}{2}\mu^{2}\phi^{a}\phi^{a}+
\frac{\lambda}{4}(\phi^{a}\phi^{a})^{2}, 
\label{lagrangean}
\end{equation}
where we notice that the mass term has a minus sign to allow a
 spontaneous 
symmetry breaking. 
The trace in the partition function is implemented by imposing the
following periodic boundary conditions on the bosonic fields
\begin{equation}
\phi^{a}(\tau,{\bf x})=\phi^{a}(\tau+\beta,{\bf x}).  
\label{bc}
\end{equation}
Because of this periodicity we may perform a Fourier expansion
for the fields in the $\tau$ variable by writting
\begin{equation}
\phi^{a}(\tau,{\bf x})=\beta^{-\frac{1}{2}}\sum_{n=-\infty}^{+\infty}
\phi_{n}^{a}({\bf x})e^{i\omega_{n}\tau},
\label{Fourier}
\end{equation}
so that, after integration over $\tau$ we have
\begin{equation}
S_{E}=\int d^{3}{\bf x} 
\left\{ 
\frac{1}{2} \sum_{n}
(\partial_{\mu}\phi_{n}^{a}\partial_{\mu}\phi_{-n}^{a}-
\mu^{2}\phi_{n}^{a}\phi_{-n}^{a})+
\frac{\lambda T}{4}
\sum_{n_{1},n_{2},n_{3},n_{4}}\delta(n_{1}+n_{2}+n_{3}+n_{4})
\phi_{n_{1}}^{a}\phi_{n_{2}}^{a}\phi_{n_{3}}^{b}\phi_{n_{4}}^{b} 
\right\}, 
\label{action-Fourier}
\end{equation}
where we have made use the well known relation
\begin{equation}
\int_{0}^{\beta}d\tau e^{i(\sum_{n}\omega_{n})\tau}=
\beta \delta (\sum_{n} \omega_{n}).
\end{equation}
Before explicitly computing the effective potential for this
action we can already see that the parameters will have a 
dependence on the temperature. Here the temperature enters via the
coupling $\lambda T$, but we still have to take into account a
 contribution comming from the frequencies $\omega_{n}$ in the
 computation of non-static Feynman loops. In the next section
 we explicity obtain the one-loop effective potential of the
 reduced model.

\section{The effective potential of the reduced model.}

As we have pointed out in the introduction, the effective 
three-dimensional
field theory is obtained from the integration of the
 $\omega_{n}\neq 0$ modes
in the Euclidean functional integral given by eq.(\ref{generator}).
This can be done by performing a
semi-classical expansion of $Z$ at one loop level \cite{Jackiw}. 
For this we consider the fields configurations $\phi^{a}$ around
 a particular constant field $\Phi_{0}^{a}$, the static part of 
the solution of a saddle-point
equation for the original action given by eq.(\ref{action}). 
We write 
$\phi^{a} \rightarrow \Phi_{0}^{a} + \eta^{a}$, so that, after
 performing the 
Gaussian integration over $\eta$ we are lead to
\begin{equation}
Z=e^{-S_{E}(\Phi_{0}^{a})}
\det
\left[
\left.
\frac{\delta^{2} S}{\delta \phi^{a} 
\delta \phi^{b}}
\right|_{\phi=\Phi_{0}}
\right]^{-\frac{1}{2}}
.
\label{effective-generator}
\end{equation}
The determinant given by eq.(\ref{effective-generator})
can be exactly calculated and gives
the one loop effective potential in terms of $\Phi_{0}^{a}$.
From the action given by eq.(\ref{action-Fourier}) we get
\begin{equation}
\left.
\frac{\delta^{2} S}{\delta \phi^{a} \delta \phi^{b}} 
\right|_{\phi=\Phi_{0}}
=
-\sum_{n}[(\bigtriangledown^{2}-\omega_{n}^{2}+\mu^{2})-
\lambda T(\Phi_{0}^{k}\Phi_{0}^{k}\delta^{ab}+
2 \Phi_{0}^{a}\Phi_{0}^{b})].
\end{equation}
Since we have rotational invariance of the vacuum, one 
can choose a coordinate system, in the fields configuration space,
in which $\Phi_{0}^{a}=(0,...,\phi_{0})$. Now, we define
\begin{equation}
m_{i}^{2}=\left\{ \begin{array}{ll}
                    \lambda T \phi_{0}^{2} & \mbox{if $i=1,...,N-1$} \\
                  3 \lambda T \phi_{0}^{2} & \mbox{if $i=N$}
                  \end{array},
          \right.
\end{equation}
so that
\begin{equation}
\left.
\frac{\delta^{2} S}{\delta \phi^{a} \delta \phi^{b}} 
\right|_{\phi=\Phi_{0}}
=
-\sum_{n}(\bigtriangledown^{2}-\omega_{n}^{2}+\mu^{2}-m_{i}^{2}).
\end{equation}
To proceed, we make use of the relation
\begin{equation}
Tr\ln \left( -\sum_{n}(\bigtriangledown^{2}-\omega_{n}^{2}+
\mu^{2}-m_{i}^{2})\right) =
(VT)\int \frac{d^{3}{\bf k}}{(2\pi)^{3}}
\ln \left( \sum_{n}({\bf k}^{2}+\omega_{n}^{2}-\mu^{2}+m_{i}^{2})
 \right),
\label{tr-ln}
\end{equation}
where $V$ is a volume element and the integral over ${\bf k}$ can 
be done with the aid of the identity
\cite{Peskin} 
\begin{equation}
\int \frac{d^{d}{\bf k}}{(2\pi)^{d}} \ln({\bf k}^{2}+m^{2})=
-\frac{\partial}{\partial s} 
\left[
\int \frac{d^{d}{\bf k}}{(2\pi)^{d}} ({\bf k}^{2}+m^{2})^{-s}
\right]_{s=0}.
\label{Peskin-trick}
\end{equation}
In order to simplify the expressions we define the quantities
\begin{equation}
\left\{ \begin{array}{ll}
        a_{i}^{2}=\omega_{n}^{2}-\mu^{2}+m_{i}^{2}=
                  (\frac{2\pi}{\beta})^{2}(n^{2}+\alpha_{i}^{2}) \\
        \alpha_{i}^{2}=(\frac{\beta}{2\pi})^{2}(m_{i}^{2}-\mu^{2})
        \end{array}
\right. 
\end{equation}
and split the sum over $n$ as
\begin{equation}
\sum_{n}(n^{2} + \alpha_{i}^{2})^{-s}= \alpha_{i}^{-2s}+
2 \sum_{n=1}^{\infty}(n^{2} + \alpha_{i}^{2})^{-s}.
\end{equation}
The integral over ${\bf k}$, in eq.(\ref{Peskin-trick}), is done using 
the well know result from dimensional regularization,
\begin{equation}
\int \frac{d^{d}{\bf k}}{({\bf k}^{2}+a_{i}^{2})}=
\pi^{\frac{d}{2}} \frac{\Gamma(s-\frac{d}{2})}{\Gamma(s)} 
\frac{1}{(a_{i}^{2})^{s-\frac{d}{2}}}
\end{equation}
and the sum over the Matsubara frequencies is done using an analytical 
extension of the Hurwitz zeta-function \cite{Elizalde-Romeo}.
This procedure gives for the determinant the result
\begin{equation}
\ln\det
\left[
\left.
\frac{\delta^{2} S}{\delta \phi^{a} \delta \phi^{b}}
\right|_{\phi=\Phi_{0}}
\right] = 
\frac{\partial}{\partial s}
\left[
\frac{\pi^{2}}{(2\pi)^{3}} 
\frac{\alpha_{i}^{4-2s}}{\Gamma(s)}
\left(
\frac{\beta}{2\pi}
\right)^{2s-3}
      \left(
               \Gamma(s-2)+
               4\sum_{n=1}^{\infty}(\pi n \alpha_{i})^{s-2}K_{s-2}
(2 \pi n \alpha_{i})
      \right)
\right]_{s=0},
\end{equation}
where $K_{\nu}$ is the modified Bessel function \cite{Abramowitz}.

The exact one-loop renormalized effective potential for the thermal 
linear $\sigma$ model can finally
be written,  with $m_{\pi}^{2}=\lambda T \phi_{0}^{2}$ and
$m_{\sigma}^{2}=3\lambda T \phi_{0}^{2}$ as
\begin{eqnarray}
V_{\mbox{eff}}(\phi_{0}^{2}) & = & -\frac{1}{2}\mu^{2}\phi_{0}^{2}
                                   +\frac{\lambda T}{4}\phi_{0}^{4} +
\nonumber \\
                             &   &
\beta \frac{\Gamma(-2)}{2^{5}\pi^{2}}
      \left[
      (N-1)(m_{\pi}^{2}-\mu^{2})^{2}+
           (m_{\sigma}^{2}-\mu^{2})^{2}
      \right]
- \nonumber \\ 
                             &   & \frac{\beta^{3}}{3} 
\frac{\Psi(0)}{2\pi^{2}}
      \left[
      (N-1)(m_{\pi}^{2}-\mu^{2})^{2} \int_{1}^{\infty} 
                       \frac{(t^{2}-1)^{\frac{3}{2}}}
             {e^{\beta (m_{\pi}^{2}-\mu^{2})^{\frac{1}{2}} t}-1} dt
      \right.
+ \nonumber \\
                             &   & \left. (m_{\sigma}^{2}-\mu^{2})^{2}
 \int_{1}^{\infty} 
                                   \frac{(t^{2}-1)^{\frac{3}{2}}}
                                   {e^{\beta (m_{\sigma}^{2}-\mu^{2})^
                                            {\frac{1}{2}} t}-1} dt
      \right]+ \nonumber \\
                             &   &
\frac{1}{2}\delta\mu^{2}\phi_{0}^{2}+
\frac{\delta\lambda T}{4}\phi_{0}^{4},
\label{exact-eff-pot}
\end{eqnarray}
where we have used the representation \cite{Abramowitz}
\begin{equation}
K_{\nu}(z)=
\frac{\pi^{\frac{1}{2}}(\frac{1}{2}z)^{\nu}}{\Gamma(\nu+\frac{1}{2})}
\int_{1}^{\infty} e^{-zt}(t^{2}-1)^{\nu-\frac{1}{2}} dt,
\end{equation}
for the modified Bessel function $K_{\nu}(z)$.

At this point it would be natural to start the search for the
 counterterms that suppress the ultraviolet divergences  
and make the theory finite when the limit of high-temperature is taken. 
However we know that it is more intuitive to read these quantities 
directly from the series expansion of the 
eq.(\ref{tr-ln}) in powers of $\phi_{0}^{2}$. In this case, 
instead of using eq.(\ref{Peskin-trick}) we write
\begin{equation}
\ln\det
\left[
\left.
\frac{\delta^{2} S}{\delta \phi^{a} \delta \phi^{b}}
\right|_{\phi=\Phi_{0}}
\right]
=
\sum_{s=1}^{\infty}\frac{(-1)^{s+1}}{s}(m_{i}^{2})^{s}
\int \frac{d^{3}{\bf k}}{(2\pi)^{3}} 
\sum_{n} 
\frac{1}{ \left( {\bf k}^{2}+\omega_{n}^{2}-\mu^{2} \right)^{s}},
\end{equation}
and, by defining the quantity
\begin{equation}
\alpha^{2}=(\frac{\beta}{2\pi})^{2}(-\mu^{2}),
\end{equation}
we have, after integration over ${\bf k}$ and summation over $n$, the 
expression
\begin{eqnarray}
V_{\mbox{eff}}(\phi_{0}) & = &
-\frac{1}{2}\mu^{2}\phi_{0}^{2}
+\frac{\lambda T}{4}\phi_{0}^{4}+ \nonumber \\
                         &   &
\sum_{s=1}^{\infty} \frac{(-1)^{s+1}}{s} 
\frac{1}{2^{3}\pi^{\frac{3}{2}}}
\left[
         (N-1)m_{\pi}^{2s}+m_{\sigma}^{2s}
\right]
(\frac{\beta}{2\pi})^{2s-3} \nonumber \\ 
                         &   & \left\{
          \frac{\pi^{\frac{1}{2}}}{2\alpha^{2s-4}} \frac{1}{\Gamma(s)}
          \left [
          \Gamma(s-2)+
4\sum_{n=1}^{\infty}(\pi n \alpha)^{s-2}K_{s-2}(2\pi n \alpha)
          \right ]
\right\}+ \nonumber \\
                         &   &
\frac{1}{2}\delta\mu^{2}\phi_{0}^{2}+
\frac{\delta\lambda T}{4}\phi_{0}^{4}.
\label{approx-eff-pot}
\end{eqnarray}

As explicitly pointed out by Landsman \cite{Landsman}, the possibility
 of 
dimensionally reducing a thermal field theory, depends on whether 
we are able to find thermal counterterms that cancel the divergent 
contributions as $\beta \rightarrow 0$ comming from the integration 
of non-zero Matsubara frequencies. These temperature 
dependent counterterms should be controlled by a thermal
renormalization group and fixed by the
imposition of renormalization conditions on Feynman diagrams computed
with the aid of some regularization scheme.
In spite of regularization being purely an intermediary tool in the
 process of renormalizing a quantum field theory, we know that
 dimensional reduction is scheme dependent \cite{Appelquist-Carazone}
 \cite{Ambjorn},
being optimal in BPHZ subtractions at zero momenta and temperature 
$\beta^{-1}$. We are working with a modified
minimal subtraction scheme at zero momenta and temperature 
$\beta^{-1}$, similarly to what has been done by Farakos et al
 \cite{Farakos}.
Now we are ready to find the exact form of the couterterms and then
 to compute the thermal
renormalization group functions. Before proceeding, however, let us
carefully examine the above expression. We notice that the form of 
the divergences 
($s=1,2$) are of the same type of the ultraviolet divergences of a
 theory in four 
dimensions. 
The role of the thermal counterterms we want to choose is exactly to 
cancel out those
divergences to give a theory well behaved in the high temperature
 limit.
To find the exact form of these counterterms we have to impose 
renormalization conditions to
correlation functions of the reduced theory. As we know, the
 ultraviolet divergences of a
thermal quantum field theory are the same of the equivalent zero
 temperature model, 
that is, the difference of imposing conditions at zero temperature
 and temperature $T$ 
is a finite renormalization (which is essential for dimensional
 reduction to occur).
Keeping this in mind we impose conditions consistent with the tree
 level action 
of the effective three dimensional theory. By requiring
\begin{equation}
\left.
\frac{\partial^{2}V_{\mbox{eff}}}
{\partial \phi_{0}^{2}}(\beta,\phi_{0})
\right|_{\phi_{0}=0}
=
-\mu^{2}+\delta\mu^{2}=\mu_{R}^{2},
\label{mass-cond}
\end{equation}
and
\begin{equation}
\left.
\frac{\partial^{4}V_{\mbox{eff}}}
{\partial \phi_{0}^{4}}(\beta,\phi_{0})
\right|_{\phi_{0}=0}
=
6(\lambda +\delta\lambda)T=6\lambda_{R}T,
\label{coup-cond}
\end{equation}
we get
\begin{equation}
\delta\mu^{2}=(N+2)
\left[
   \lambda \frac{T^{2}}{24}-
   \lambda \frac{(-\mu^{2})}{16\pi^{2}}
   \left[
   \frac{1}{\varepsilon}+\Psi(2)-4\sum_{n=1}^{\infty}
   \ln {(\frac{n\beta}{2}(-\mu^{2})^{\frac{1}{2}})}+
\gamma \right] \right],
\label{mass-countterm}
\end{equation}
and
\begin{equation}
\delta\lambda=-\frac{(N+8)}{6}
\left[
   \frac{\lambda^{2}}{64\pi^{2}}
   \left[
   \frac{1}{\varepsilon}+
   \Psi(1)-
   4\sum_{n=1}^{\infty}
   \ln{(\frac{n\beta}{2}(-\mu^{2})^{\frac{1}{2}})}+\gamma
  \right]
\right],
\label{coup-countterm}
\end{equation}
where $\Psi(z)$ is the usual Psi function and $\gamma$ is the Euler
 number
$\Psi(1)=-\gamma$ \cite{Abramowitz}.
Choosing those quantities as mass and coupling constant counterterms
 we guarantee that
the new effective action for the reduced theory is finite for very 
high, but still finite, temperatures. However in the effective 
potential given by eq.(\ref{approx-eff-pot}), there is a new 
contribution,
 independent of $T$, coming 
from the evaluation of the six-point function. The computation of
 the
six-point function with non-static internal loops gives a contribution
for the static effective potential, as expected since we do not have 
obtained a
perfect decoupling of non-static modes, but also gives an extra 
interaction term which was not present in the original action 

\begin{equation}
\left.
\frac{\partial^{6}V_{\mbox{eff}}}{\partial\phi_{0}^{6}}(\beta,\phi_{0})
\right|_{\phi_{0}=0}
=...+
\frac{(N+26)}{9}\frac{\lambda^{3}\zeta(3)}{2^{9}\pi^{4}},
\label{phi-6-term}
\end{equation}
exactly in the same sense as in the case of the scalar $\phi^{4}$ 
theory
\cite{Landsman}.
We also see that, as expected, the non-renormalizable interaction terms
 are suppressed 
by negative powers of $T$ in the limit of high temperatures. Indeed, 
for $s>2$ the sum
over $n$ is analytic and is expressed in terms of usual zeta functions. 
Hence we
are left, as we can see from eq.(\ref{approx-eff-pot}), with
$T^{s}(\frac{\beta}{2\pi})^{2s-3} \sim T^{3-s}=T^{3-\frac{N}{2}}$ as
coeficient of the sum over $n$. From this behavior we can check the
temperature dependence of the $2$, $4$ and $6$-point functions computed
and neglect all the $N>6$ contributions.

Finally, the effective three dimensional reduced theory has the 
following action
\begin{equation}
\Gamma(\phi_{0})=
\int d^{3}{\bf x}
\left\{
          \frac{1}{2}(\partial_{i}\phi_{0})^{2}+ 
          \frac{1}{2}\mu_{R}^{2}\phi_{0}^{2}+
          \frac{\lambda_{R} T}{4}\phi_{0}^{4}+
          \frac{(N+26)}{9}
\frac{\lambda_{R}^{3}\zeta(3)}{2^{9}\pi^{4}}
\phi_{0}^{6},
\right\}.
\label{reduced-action}
\end{equation}
We see from eq.(\ref{reduced-action}) that, as long as $T$ is finite, 
the reduced model defines a well behaved theory at the tree level.
 In the 
limit of infinite temperatures the "new" bare coupling constant
 $\lambda_{R}T$ 
imposes a second "renormalization" procedure. This action 
has the form of the standard action
for a massive scalar field with $\varphi^{4}$ and $\varphi^{6}$
 self-interaction terms
\begin{equation}
\Gamma(\varphi)=\int d^{3}{\bf x}
\left(
           \frac{1}{2}(\partial_{i} \varphi)^{2}+
           \frac{a}{2}\varphi^{2}+
           \frac{b}{4}\varphi^{4}+
           \frac{c}{6}\varphi^{6}
\right),
\label{action-phi-6}
\end{equation}
in a three-dimensional space. The study of the criticality of such a
 system
can be found in a great variety of text books (see for instance 
\cite{Zinn-Justin})
and a general discussion can help us to understand the consequences
of dimensionally reducing a quantum field theory which has a
 symmetry-restoring
phase, as in our case. In the study of mean-field theory for
 ferromagnetic systems of the type given by eq.
(\ref{action-phi-6}) we have situations in which the parameters
 $a$ and $b$ (c
is fixed in order to have a potential bounded from below)
 can be made to vanish.
This occurs for instance in $He^{3}-He^{4}$ mixtures or some 
metamagnetic systems.
Depending on the values of $a$ and $b$, we can have first-order
 phase transitions, second-order phase transitions or both 
(the tricritical phenomenon \cite{Gino}). If we carefully 
examine eq.(\ref{reduced-action}) we see that
one of the relevant parameters, the renormalized thermal coupling
 $\lambda_{R}$ defined in eq.(\ref{coup-cond}), is positive, at least
 for small $N$, since the bare quantity $\lambda$ is also positive 
by definition. Thus
we do not have a change in sign for both $\varphi^{4}$ and 
$\varphi^{6}$ terms in eq.(\ref{action-phi-6}) and the reduced theory 
has a stable vacuum. On the other hand, if we look for the mass term 
in the 
reduced action of eq.(\ref{reduced-action}) we see that in the limit of 
high temperatures, its thermal correction causes a change in sign and
 the final expression is exactly the one in eq.(\ref{action-phi-6}) with
 all positive 
coeficients, i.e. we do not have the possibility of phase transitions
 of any kind. The conclusion is that in the limit of high temperatures
the symmetry-restoring phase of the former model is unambigously mapped
 in the reduced one. Finally, to consider all the possibilities, we
 still have to argue what happens in the limit $N \rightarrow \infty$.
 In this case, the renormalized
coupling $\lambda_{R}$ has its sign changed for some suficiently large
 $N$, as we can see from eq.(\ref{coup-countterm}), and, in this limit, 
the reduced model is the one of a unstable theory. This result has
 been obtained by 
Linde \cite{Linde} and more recently by Da Silva \cite{Da Silva}.

For completeness we finally compute the thermal renormalization group 
functions. As previously said in the introduction, the transition from
 a four-dimensional
theory to an effective three-dimensional one is renormalization point
 dependent.
This is not surprising since the Appelquist-Carazone decoupling theorem 
only holds if a 
particular class of renormalization prescriptions is adopted. 
Neverthless we
can write down renormalization group equations. The independence 
of bare correlation functions on the choice of the renormalization
 point, expressed by the conditions
\begin{eqnarray}
\mu\frac{d}{d\mu}\Gamma^{(N)}=0, \\
T\frac{d}{dT}\Gamma^{(N)}=0,
\end{eqnarray}
give us those equations.
Since we are only interested on the computation of thermal 
renormalization group
functions, we will consider solely

\begin{equation}
\left(
         T\frac{\partial}{\partial T}+
         \beta_{T}\frac{\partial}{\partial\lambda}+
         \gamma_{T}m^{2}\frac{\partial}{\partial m^{2}}
\right)
\Gamma^{(N)}=0
\label{rg-equation}
\end{equation}
where
\begin{equation}
\beta_{T}=T\frac{\partial\lambda}{\partial T},
\end{equation}
\begin{equation}
\gamma_{T}=m^{-2}T\frac{\partial m^{2}}{\partial T},
\end{equation}
and
\begin{equation}
\Gamma^{(N)}=
\left.
\frac{\partial^{N}V_{\mbox{eff}}}{\partial\phi_{0}^{N}}
\right|_{\phi_{0}=0}.
\end{equation}
Using the results of the previous section we easily get
\begin{equation}
\beta_{T}=\frac{(N+8)}{6}\frac{\lambda^{2}}{16 \pi^{2}}
\label{beta-function}
\end{equation}
and 
\begin{equation}
\gamma_{T}=(N+2)
           \left[ \frac{\lambda T^{2}}{12 m^{2}}+
                     \frac{\lambda}{16 \pi^{2}} \right],
\label{gama-function}
\end{equation}
as reasonably expected.
If we now solve the eqs.(\ref{beta-function}) and 
(\ref{gama-function}) we obtain the renormalized parameters
$m$ and $\lambda$ as functions of $T$, i.e. the thermal renormalized
running mass and coupling constant.

\section{Conclusions}

In this paper the dimensional reduction of the $O(N)$ invariant linear
 $\sigma$ model is performed. The effective potential for the reduced 
theory is exhibited in two different ways. From its exact form we can
 see that the ultraviolet divergences for the reduced theory are of 
the same kind of the ones for the original theory. On the other side,
 from its series expansion on the static fields we easily obtain the
 thermal counterterms that allow dimensional reduction to take place.
 Moreover, a new renormalizable interaction term, for the reduced 
theory, came up without changing the critical behavior of the former 
one. Some of the main features of dimensional reduction are present,
 as the fact that the effective action is a polinomial on the field
 because nonrenormalizable interaction terms are suppressed by powers
 of $\beta$ in the limit of high temperatures ($\beta\rightarrow 0$).
 In addition, we also find, in the reduced model, a characteristic 
already observed by Linde \cite{Linde}, in another context, in which 
the potential becomes unstable for large values of $N$. A natural 
continuation of this paper would be to perform the dimensional
 reduction
 of the non-linear $\sigma$ model. It also might be interesting to
 ask what happens to the tree-dimensional eletrodynamics with a 
Maxwell-Chern-Simons term in the high-temperature regime after 
integration over both fermionic and non-zero bosonic Matsubara 
modes. Note that the inclusion, in the Lagrangian, of the Chern-Simons
 term avoids infrared divergences. These subjects are already under 
investigation.  

\section{Acknowledgment}

This paper was
suported by Conselho Nacional de 
Desenvolvimento Cientifico e Tecnologico do Brazil (CNPq) and 
Centro de Apoio a Pesquisa (CAPES).

\end {document}